\documentclass[dvips]{acta}
\usepackage{supertabular,lscape,epsfig}
\usepackage{amssymb}
\usepackage{amsmath}
\SetPages{1}{9}
\SetVol{63}{2013}

\begin{document}

\begin{Titlepage}

\Title { On the Nature of Superhumps }

\Author {J.~~S m a k}
{N. Copernicus Astronomical Center, Polish Academy of Sciences,\\
Bartycka 18, 00-716 Warsaw, Poland\\
e-mail: jis@camk.edu.pl }

\Received{  }

\end{Titlepage}

\Abstract { Further evidence is presented supporting the alternative 
interpretation of superhumps as being due to irradiation modulated 
periodically variable mass transfer rate. 
NZ Boo, HT Cas and PU UMa are added to the sample of high 
inclination dwarf novae showing -- during their superoutbursts -- 
modulation of the observed brightness of the disk with beat period. 
Simple model calculations confirm earlier hypothesis that this modulation 
is due to a non-axisymmetric structure of the outer parts of the disk, 
involving the azimuthal dependence of their geometrical thickness, 
rotating with the beat period. 

The modulation amplitude $A_{mod}$ is found to decrease during superoutbursts.  
In particular, it is found that during two superoutbursts of OY Car the rate 
of decline of the superhump amplitude $dA_{SH}/dt$ was correlated with 
the rate of decline $dA_{mod}/dt$. 
This leads to a simple explanation for the decreasing amplitudes of 
superhumps: it is a consequence of the decreasing modulation amplitude. 
} 
{accretion, accretion disks -- binaries: cataclysmic variables, 
stars: dwarf novae, stars: individual: NZ Boo, OY Car, HT Cas, DV UMa, PU UMa. }

\section { Introduction } 

Superhumps are periodic light variations observed in dwarf novae of the SU UMa 
subtype during their superoutbursts (Warner 2003); they are also observed in 
the so-called permanent superhumpers (cf. Patterson 1999). 
The superhump periods are slightly longer than the orbital periods and show 
complex variations (Kato et al. 2009, 2012 and references therein). 
Their amplitudes at superoutburst maximum are from 0.2 to 0.7 mag. and decrease 
afterwards at a rate from $dA_{SH}/dt=-0.02$ to $-0.05$ mag/day (Smak 2010 
and references therein). 

The commonly accepted tidal-resonance model, first proposed by Whitehurst (1988) 
and Hirose and Osaki (1990), explains superhumps as being due 
to tidal effects in the outer parts of accretion disks, leading -- via the 3:1 
resonance -- to the formation of an eccentric outer ring undergoing apsidal motion. 
This model and, in particular, the results of numerous 2D and 3D SPH 
simulations (cf. Smith et al. 2007 and references therein) successfully reproduce 
the observed superhump periods; this suggests that the basic "clock" which 
defines the superhump periods may indeed be related to the apsidal motion. 

On the other hand, however, the tidal-resonance model fails to explain many 
other important facts (cf. Smak 2010). In particular it fails to reproduce 
the amplitudes of superhumps: The numerical SPH simulations produce "superhumps" 
with amplitudes which -- compared to the observed amplitudes  -- are about 10 
times too low (Smak 2009a). 

From the analysis of superoutburst light curves of several dwarf novae it was 
found (Smak 2009b) that in the case of deeply eclipsing systems ($i>82^\circ$) 
the observed brightness of the disk is modulated with phase of the beat period. 
This was interpreted as being due to a non-axisymmetric structure of the outer 
parts of the disk, involving azimuthal dependence of their geometrical thickness, 
rotating with the beat (apsidal motion?) period. 

Regardless of this interpretation it is obvious that the modulation with 
the beat period seen by the observer implies modulated irradiation of 
the secondary component with the superhump period. 
This became one of the crucial ingredients of the new interpretation of 
superhumps (Smak 2009b) as being due to periodically variable dissipation 
of the kinetic energy of stream resulting from irradiation controlled 
mass outflow from the secondary. 

The purpose of this paper is to present new results which strengthen this 
interpretation. The original sample of deeply eclipsing systems showing 
modulation effects described above is enlarged by adding three new objects 
(Section 2). The interpretation of the observed modulation in terms of 
the non-axisymmetric structure of rotating disk is confirmed (Section 3). 
And -- most importantly -- the modulation amplitude is shown to decrease during 
superoutbursts (Section 4) providing natural explanation for the decreasing 
amplitudes of superhumps.

\section { Modulation with Beat Period: New Data } 

A search through the recent literature produced three high inclination systems 
with well covered superoutburst light curves: NZ Boo = J1502+3334 
(Shears et al. 2011), HT Cas (Kato et al. 2012) and PU UMa = J0901+4809 
(Shears et al. 2012). 
From those light curves the magnitudes $m_d$ of the disk were extracted 
by using phases away from superhump maxima and away from eclipses.  
The values of $m_d$ were fitted with a simple formula:  

\begin{equation}
m_d~=~m~(t_\circ)~+~{{dm}\over{dt}}~(t-t_\circ)~+~\Delta m~,  
\end{equation} 

\noindent
or, in the case of non-linear variations, 

\begin{equation}
m_d~=~m~(t_\circ)~+~{{dm}\over{dt}}~(t-t_\circ)~+~{{d^2m}\over{dt^2}}~
     (t-t_\circ)^2~+~\Delta m~, 
\end{equation} 

\noindent
where the first two (or three) terms describe the slow brightness decline 
during superoutburst, while the last one -- the expected modulation with 
the beat period: 

\begin{equation}
\Delta m~=~-~A_{mod}~\cos~(\phi_b-\phi_{b,max})~. 
\end{equation} 

Results are shown in Fig.1 and listed in Table 1. Listed in this table are 
also all earlier determinations for systems with $i\ge 80^\circ$. 
Orbital inclinations given in the second column are from the latest edition 
of the Catalogue of CV's by Ritter and Kolb (2003); the inclination of PU UMa 
is unknown. 
As can be seen the expected modulation with beat phase is clearly present 
in all three systems. HT Cas and PU UMa are very similar to OY Car, DV UMa, 
IY UMa and J1227 studied earlier. In particular, their phases of maximum fall 
in a narrow interval from $\phi_{b,max}=0.55$ to 0.68. 
NZ Boo, however, turns out to be an exception: its curve is shifted with 
respect to the others by -- roughly -- 0.5. 

\vskip -5truemm
\begin{figure}[htb]
\epsfysize= 8.5cm 
\hspace{3.5cm}
\epsfbox{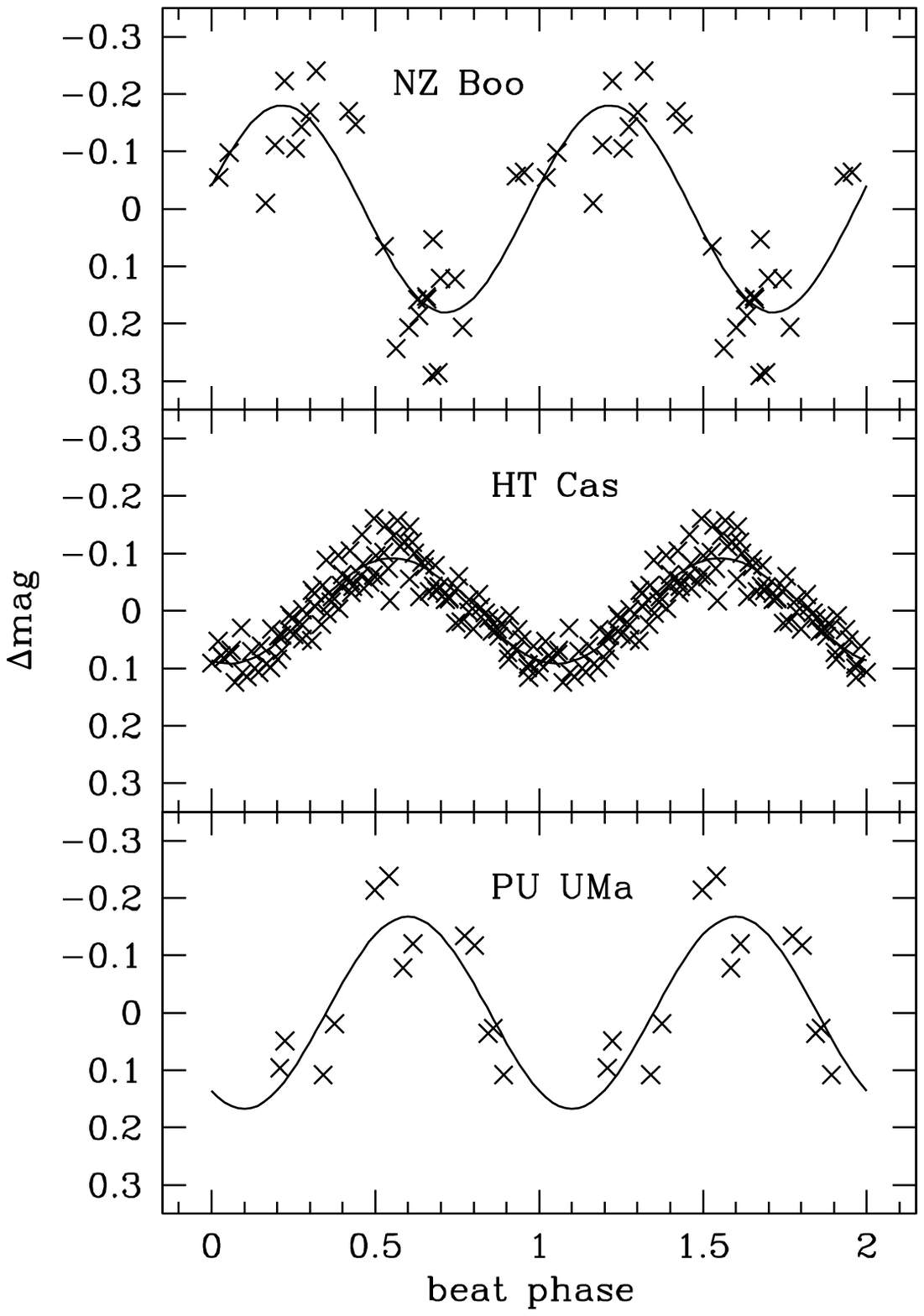} 
\vskip -5truemm
\FigCap { Residual magnitudes {\it vs.} beat phase. Solid lines are cosine 
curves with $A_{mod}$ and $\phi_{b,max}$ listed in Table 1. }
\end{figure}

\begin{table}[h!]
{\parskip=0truept
\baselineskip=0pt {
\medskip
\centerline{Table 1}
\medskip
\centerline{ Mean Modulation Parameters }
\medskip
$$\offinterlineskip \tabskip=0pt
\vbox {\halign {\strut
\vrule width 0.5truemm #&	
\enskip\hfil#\enskip&	        
\vrule#&			
\enskip#\enskip&                
\vrule#&			
\enskip#\enskip&	        
\vrule#&			
\enskip#\enskip& 	        
\vrule#&			
\enskip#\hfil\enskip&   	
\vrule width 0.5 truemm # \cr	
\noalign {\hrule height 0.5truemm}
&&&&&&&&&&\cr
&\hfil Star\hfil&&\hfil $i$\hfil&&\hfil $A_{mod}$ (mag)\hfil&&\hfil$\phi_{b,max}$ 
     \hfil&&Notes&\cr
&&&&&&&&&&\cr
\noalign {\hrule height 0.5truemm}
&&&&&&&&&&\cr
&   NZ Boo && 88.4 $\pm$ 0.2 &&  0.18 $ \pm$ 0.03 && 0.23 $\pm$ 0.03 && 4,5 &\cr
&   OY Car && 83.3 $\pm$ 0.2 &&  0.20 $ \pm$ 0.07 && 0.65 $\pm$ 0.07 && 1   &\cr
&   HT Cas && 81.0 $\pm$ 1.0 &&  0.10 $ \pm$ 0.01 && 0.55 $\pm$ 0.01 && 4,6 &\cr
&    Z Cha && 81.8 $\pm$ 0.1 &&  0.06 $ \pm$ 0.09 && 0.28 $\pm$ 0.25 && 1   &\cr
&   XZ Eri && 80.0 $\pm$ 0.1 &&  0.00 \hfil       &&                 && 1,3 &\cr
&   DV UMa && 82.9 $\pm$ 0.1 &&  0.17 $ \pm$ 0.06 && 0.60 $\pm$ 0.06 && 1   &\cr
&   IY UMa && 86.0 $\pm$ 1.0 &&  0.18 $ \pm$ 0.04 && 0.67 $\pm$ 0.04 && 1   &\cr
&   PU UMa &&                &&  0.17 $ \pm$ 0.06 && 0.60 $\pm$ 0.05 && 4,7 &\cr
&J1227+5139&& 84.3$ \pm$ 0.1 &&  0.18 $ \pm$ 0.03 && 0.68 $\pm$ 0.03 && 2   &\cr
&&&&&&&&&&\cr
\noalign {\hrule height 0.5truemm}
}}$$
\parindent=0 truemm
\parskip=3truept
\leftskip=30truept
\rightskip=30truept

Notes: 1. Smak (2009b); 2. Smak (2010); 3. Uemura et al. (2004); 
4. this paper; 5. data from Shears et al. (2011), Figs.2 and 3 (see Appendix A); 
6. data from Kato et al. (2012), Figs.9 and 11(a); 
7. data from Shears et al. (2012), Fig.2. 

\bigskip
\parindent=12 truemm
\parskip=12truept
\leftskip=0pt
\rightskip=0pt
}}
\end{table}

Fig.2 shows the full amplitudes (i.e. $2A_{mod}$) of systems listed in Table 1 
plotted against their orbital inclination. In the case of NZ Boo, taking into 
account its $\phi_{b,max}$, the amplitude is plotted with {\it minus} sign 
(the same is done for Z Cha, with $\phi_{b,max}=0.28\pm 0.25$, although 
in this case the amplitude is practically equal to zero). 
HT Cas is plotted with $i=81^\circ$ listed in the latest edition of 
the Catalogue of CV's by Ritter and Kolb (2003). As discussed in Appendix B, 
however, this value is uncertain, another determination giving $i=83^\circ$ 
(shown by arrow in Fig.2). At this point we note that (1) large amplitudes 
occur at inclinations $i\sim 83-86^\circ$, and (2) NZ Boo, with its inclination 
much closer to $90^\circ$, looks peculiar.

\begin{figure}[htb]
\epsfysize= 10.0cm 
\hspace{2.0cm}
\epsfbox{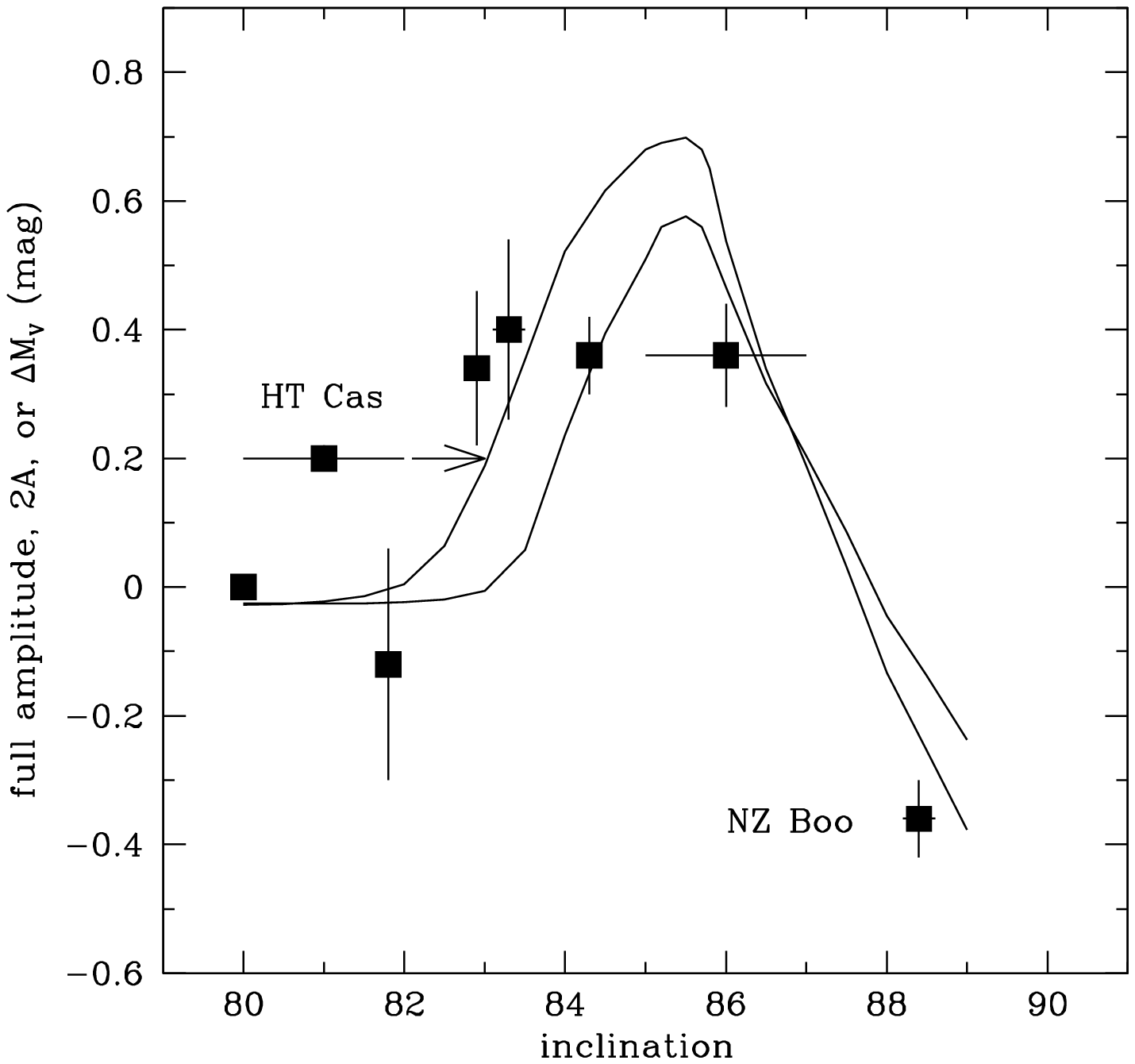} 
\vskip -25truemm
\FigCap { Full modulation amplitude $2A_{mod}$ {\it vs.} orbital inclination 
for stars listed in Table 1. An arrow shows alternative value of the inclination 
for HT Cas (see text and Appendix B). Solid lines are model curves corresponding 
to 40 and 60 percent variations in $z/r$ of the outer disk edge.  }
\end{figure}

\section { Interpretation } 

As mentioned in the Introduction it was suggested (Smak 2009b) that the observed 
modulation with beat period could be due to a non-axisymmetric structure of 
the outer parts of the disk, involving the azimuthal dependence of their 
geometrical thickness, rotating with the beat period. 
In this Section we use simple calculations to confirm this explanation.    

To begin with we calculate the "standard" absolute visual magnitudes 
$M_V^\circ(i)$ of an accretion disk as seen at a given orbital inclination. 
The details are as follows. 
The distribution of temperature on the surface of the disk is obtained from 
the standard formula 

\begin{equation}
\sigma T_e^4~=~{3\over{8\pi}}~{{GM_1}\over {R^3}}~\dot M~
		 [~1~-~(R_1/R)^{1/2}~]~.
\end{equation}

\noindent
With those temperatures we then calculate the local fluxes (using calibration 
based on Kurucz [1993] model atmospheres) and the local values of $z/r$ 
(Smak 1992) which determine the concave shape of the disk.  
Turning to the disk edge -- its vertical extent is defined by 
$(z/r)_{edge}=(z/r)_{disk}$ at $R=R_{disk}$ and its temperature by 
$T_{e,edge}=T_{e,disk}$ also at $R=R_{disk}$. 

In the next step we repeat those calculations with 

\begin{equation}
(z/r)_{edge}~=~f~(z/r)_{edge}^\circ~,    
\end{equation}

\noindent
where $(z/r)_{edge}^\circ$ is the standard value used earlier and factor $f$ 
accounts for the azimuthal variations of the vertical extent of the disk edge. 
The resulting magnitudes $M_V(i,f)$ are then used to obtain the modulation 
amplitude 

\begin{equation}
\Delta M_V(i,f)~=~M_V(i,f)~-~M_V^\circ(i)~.  
\end{equation}

Initial test calculations, using different sets of system parameters, 
showed that the resulting values of $\Delta M_V(i,f)$ depend only weakly on 
the particular choice of those parameters. 
Final calculations were then made using system parameters of Z Cha with 
$\dot M=2\times 10^{17}$ g/s. 

Results, for $f=1.4$ and 1.6, are shown in Fig.2. Their interpretation is 
quite simple. 
At inclinations $i<82^\circ$ the dominant contributions to the observed 
luminosity comes from the surface of the disk, its obscuration by the outer 
edge is negligible, hence no modulation is observed. 
At inclinations $82^\circ <i<86^\circ$ the surface of the disk is partly 
obscured by the disk edge; when the disk edge is more extended ($f>1$) 
this obscuration becomes larger making $M_V(i,f)$ fainter. 
At inclinations $i>87^\circ$, however, the main contribution to the observed 
luminosity comes from disk edge; when it is more extended ($f>1$) its area 
is larger making $M_V(i,f)$ brighter. 

Finally, we compare our results with modulation amplitudes observed in systems 
listed in Table 1. As can be seen from Fig.2, the location of the observed points 
is reasonably well represented by the two curves corresponding to 40-60 percent 
asymmetry in the vertical extent of disk edge.

\section { Decreasing Modulation Amplitude } 

An inspection of superoutburst light curves of OY Car (Fig.1a in Krzemi{\'n}ski 
and Vogt 1985 and Fig.1 in Schoembs 1986) and HT Cas (Figs.11(a) and 12(b) in 
Kato {\it et al.} 2012) shows that the modulation amplitude $A_{mod}$ {\it decreases} 
during superoutburst. 

To study this effect in detail we need light curves covering several beat cycles. 
Sufficient data meeting this requirement were found for only six cases listed 
in Table 2 (inluding two different superoutbursts of OY Car). 
We analyze them as before (Section 2) with Eq.(3) being now replaced with   

\begin{equation}
\Delta m~=~-~\left[~A_{mod,\circ}~+~{{dA_{mod}}\over{dt}}~(t-t_\circ)\right]~
         \cos~(\phi_b-\phi_{b,max})~, 
\end{equation} 

\noindent
where $dA_{mod}/dt$ describes variations of the modulation amplitude during 
superoutburst.

\begin{table}[h!]
{\parskip=0truept
\baselineskip=0pt {
\medskip
\centerline{Table 2}
\medskip
\centerline{ Variable Modulation Parameters }
\medskip
$$\offinterlineskip \tabskip=0pt
\vbox {\halign {\strut
\vrule width 0.5truemm #&	
\enskip#\enskip&	        
\vrule#&			
\enskip\hfil#\hfil\enskip&	
\vrule#&			
\enskip\hfil#\hfil\enskip&	
\vrule#&			
\enskip#\hfil\enskip&	        
\vrule width 0.5 truemm # \cr	
\noalign {\hrule height 0.5truemm}
&&&&&&&&\cr
&\hfil Star\hfil&&$A_{mod,\circ}$(mag)&&$dA_{mod}/dt$(mag/day)&& Notes &\cr
&&&&&&&&\cr
\noalign {\hrule height 0.5truemm}
&&&&&&&&\cr
& NZ Boo\hfil && 0.18 $\pm$ 0.05 && $-$ 0.002 $\pm$ 0.015 && 1 &\cr
& OY Car,\hfil Jan.1980&& 0.33 $\pm$ 0.03 && $-$ 0.039 $\pm$ 0.007 && 2 &\cr
& OY Car, Nov.1980&& 0.47 $\pm$ 0.04 && $-$ 0.086 $\pm$ 0.010 && 3 &\cr
& HT Cas\hfil && 0.13 $\pm$ 0.01 && $-$ 0.007 $\pm$ 0.002 && 4 &\cr
& DV UMa\hfil && 0.16 $\pm$ 0.08 && $+$ 0.003 $\pm$ 0.011 && 5 &\cr
& PU UMa\hfil && 0.42 $\pm$ 0.03 && $-$ 0.048 $\pm$ 0.005 && 6 &\cr
&&&&&&&&\cr
\noalign {\hrule height 0.5truemm}
}}$$
\parindent=0 truemm
\parskip=3truept
\leftskip=30truept
\rightskip=30truept

Data sources: 1. Shears et al. (2011), Figs.2 and 3 (see Appendix A); 
2. Krzemi{\'n}ski and Vogt (1985), Fig.2b; 3. Schoembs (1986), Fig.1; 
4. Kato et al. (2012), Figs.9 and 11(a); 5. Patterson et al. (2000), Fig.2; 
6. Shears et al. (2012), Fig.2.

\bigskip
\parindent=12 truemm
\parskip=12truept
\leftskip=0pt
\rightskip=0pt
}}
\end{table}

Results, listed in Table 2, show that in four cases $dA_{mod}/dt<0$, 
with individual values being significant at a $3\sigma-9\sigma$ level, 
while in the remaining two cases -- NZ Boo and DV UMa -- $dA_{mod}/dt\approx 0$. 
In fact, however, the values of $dA_{mod}/dt$ for those two stars are 
quite uncertain: compared to the other four cases their errors are twice 
s large and it is obvious that $dA_{mod}/dt<0$ cannot be excluded. 

Taking this into account we conclude that {\it the modulation amplitudes 
decrease with time during superoutburst}. 
Within our interpretation of superhumps, in terms of irradiation modulated 
periodically variable mass transfer rate, this provides -- for the first time -- 
a simple explanation for the decreasing amplitudes of superhumps: 
it is a consequence of the decreasing modulation amplitude. 

As emphasized earlier (Smak 2009b), our interpretation of superhumps is based  
on purely observational evidence and no model is available at present which 
would permit to express the observed superhump amplitude as a function of 
the observed modulation amplitude (and system parameters!). 
In spite of that, however, it is possible to test the $A_{SH}-A_{mod}$ 
connection by using data available for the two superoutbursts of OY Car. 
As can be seen from Table 2, the values of $dA_{mod}/dt$ obtained from 
those two superoutbursts are significantly different. 
Adopting our interpretation we predict that -- as a consequence -- 
the values of $dA_{SH}/dt$, describing the decreasing superhump 
amplitudes during those two superoutbursts should also be different.

\begin{figure}[htb]
\epsfysize= 13.0cm 
\hspace{1.0cm}
\epsfbox{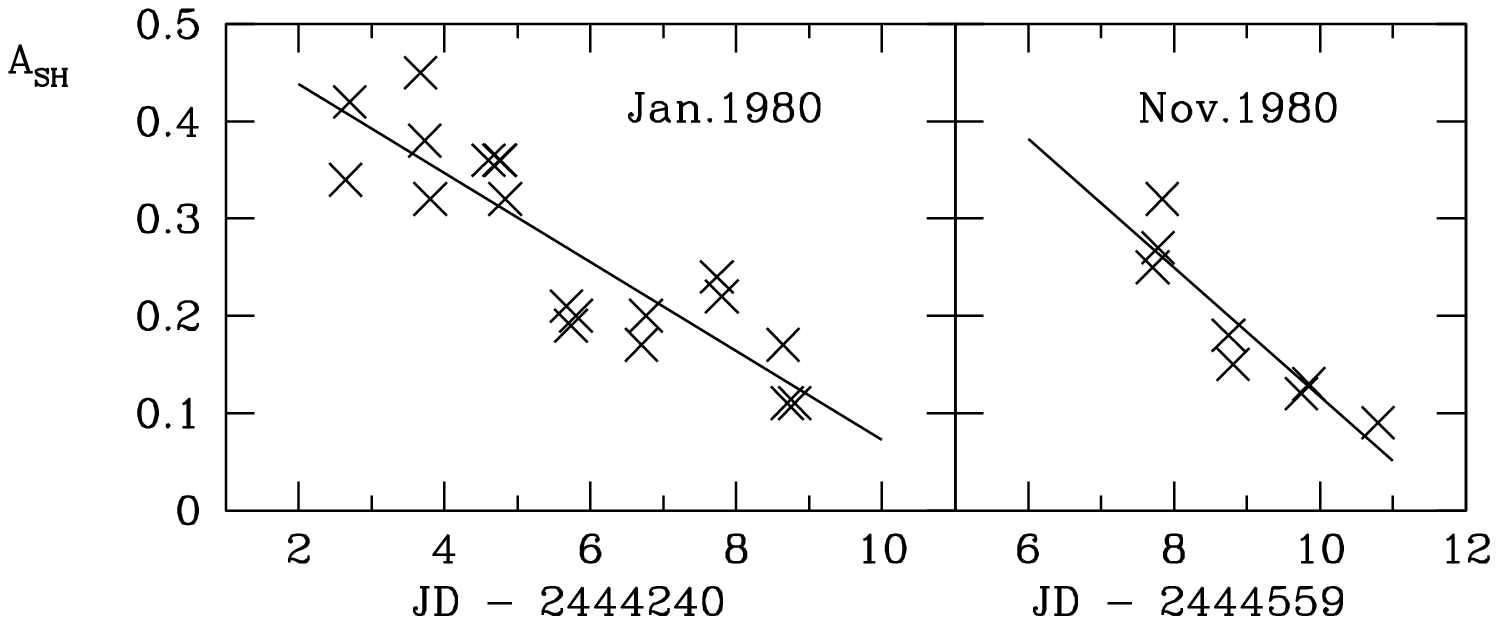} 
\vskip -85truemm
\FigCap {Superhump amplitudes {\it versus} time during two superoutbursts 
of OY Car. See text for details. }
\end{figure}

From light curves published by Krzemi{\'n}ski and Vogt (1985) 
and Schoembs (1986) we determine the superhump amplitudes $A_{SH}$; 
they are plotted in Fig.3. 
Their variations during the superoutburst are described with  

\begin{equation}
A_{SH}~=~A_{SH,\circ}~+~{{dA_{SH}}\over {dt}}~(t~-~t_\circ)~, 
\end{equation} 

\noindent
giving $dA_{SH}/dt=-0.046\pm0.006$ mag/day for the January 1980 superoutburst 
and $dA_{SH}/dt=-0.066\pm0.012$ mag/day for the November 1980 superoutburst. 
Our prediction is then confirmed: during the November 1980 superoutburst, 
when the modulation amplitude decreased faster, the superhump amplitude 
also decreased faster.

\section { Appendix A: The Light Curves of NZ Boo } 

The magnitudes of the disk, $m_d$, were determined in the following way. 
From the low resolution light curves, shown in Fig.2 of Shears et al. (2011), 
only the magnitudes corresponding to superhump maxima, $m_{max}$, 
and magnitudes corresponding to the eclipses, $m_{min}$, could be determined. 
From light curves (with arbitrary zero points) shown in their Fig.3 
the differences $\Delta m_1=m_d-m_{max}$ and $\Delta m_2=m_{min}-m_d$ 
were determined. The values of $m_d$ were then obtained as 

\begin{equation}
m_d~=~m_{max}~+~\Delta m_1~,~~~~{\rm or}~~~~m_d~=~m_{min}~-~\Delta m_2~. 
\end{equation} 

\noindent
Unfortunately it turned out that in the case of data for JD 2455022 and 023 
the full ranges, $\Delta m=m_{min}-m_{max}$, obtained from Fig.2 differ 
considerably from those obtained from Fig.3. Consequently only the light 
curves observed on JD 2455024-031 could be used in the analysis.

\section { Appendix B: On the Inclination of HT Cas }

The orbital inclinations of dwarf novae are most reliably determined 
from the simultaneous analysis of eclipses of the standard hot spot and 
of the white dwarf observed at quiescence. 
Regretfully, the light curves of HT Cas (Patterson 1981, Zhang et al 1986, 
Horne et al. 1991, Feline 2005, and others) are peculiar: only very seldom 
they show well defined hot spot eclipses (and very rarily the characteristic 
"orbital hump"). 

The inclination of HT Cas, $i=81.0\pm 1.0$, listed in the latest edition of 
the Catalogue of CV's by Ritter and Kolb (2003), comes from Horne et al. (1991). 
Their determination was based on three hot spot eclipses observed by 
Patterson (1981) for which the mean phases of ingress, $\phi_i=-0.007$, 
and egress, $\phi_e=0.072$, could be determined. Using those values they 
found the mass ratio $q=0.15$ and -- {\it via} the $i-q$ relation 
(Fig.11 in Patterson 1981) -- the inclination $i=81.0$.  
Worth adding, however, is that the mass ratio is determined primarily by the 
value of $\phi_e$ (see Fig.4a in Horne et al. 1991) and that their $\phi_e$ was 
based on only one eclipse for which the phase $\phi_4$ was measurable. 

Among the light curves of HT Cas observed by Zhang et al. (1986) there is 
another one with well defined "orbital hump" and the hot spot eclipse (E26353). 
The phases of egress obtained from this eclipse are: 
$\phi_3=0.074, \phi_4=0.090$ and $\phi_e=0.082$. 
With this value of $\phi_e$, using Fig.4a of Horne et al. (1986) we get 
$q\approx 0.10$, and -- using Fig.11 of Patterson (1981) -- $i\approx 83$. 
This shows that the inclination of HT Cas remains quite uncertain.

\begin {references} 

\refitem {Feline, W.J., Dhillon, V.S., Marsh, T.R., Watson, C.A., Littlefair, S.P.} 
         {2005} {\MNRAS} {364} {1158} 

\refitem {Hirose, M., Osaki, Y.} {1990} {\em Publ.Astr.Soc.Japan} {42} {135}

\refitem {Horne, K., Wood, J.H., Stiening, R.F.} {1991} {\ApJ} {378} {271} 

\refitem {Kato, T. {\it et al.}} {2009} {\em Publ.Astr.Soc.Japan} {61} {395}

\refitem {Kato, T. {\it et al.}} {2012} {\em Publ.Astr.Soc.Japan} {64} {21}

\refitem {Krzemi{\'n}ski, W., Vogt, N.} {1985} {\AA} {144} {124} 

\refitem {Kurucz, R.L.} {1993} {\rm CD-Rom No 13} {~} {~}

\refitem {Patterson, J.} {1981} {\ApJS} {45} {517} 

\refitem {Patterson, J.} {1999} {{\it Disk Instabilities in Close Binary Systems}, 
         Eds. S.Mineshige and J.C.Wheeler (Tokyo: Universal Academy Press)} {~} {61} 

\refitem {Patterson, J., Vanmunster, T., Skillman, D.R., Jensen, L., 
         Stull, J., Martin, B., Cook, L.M., Kemp, J., Knigge, C.}  
         {2000} {\PASP} {112} {1584} 

\refitem {Ritter, H., Kolb, U.} {2003} {\AA} {404} {301, update RKcat 7.19,2013}

\refitem {Schoembs, R.} {1986} {\AA} {158} {233} 

\refitem {Shears, J. {\it et al.}} {2011} {\it {J.Br.Astron.Assoc.}} {121} {96} 

\refitem {Shears, J., Hambsch, F.-J., Littlefield, C., Miller, I., Morelle, E., 
      Pickard, R., Pietz, J., Sabo, R.} {2012} {\rm arXiv astro-ph} {~} {1209.4062} 

\refitem {Smak, J.} {1992} {\Acta} {42} {323}  

\refitem {Smak, J.} {2009a} {\Acta} {59} {103} 

\refitem {Smak, J.} {2009b} {\Acta} {59} {121}

\refitem {Smak, J.} {2010} {\Acta} {60} {357} 

\refitem {Smith, A.J., Haswell, C.A., Murray, J.R., Truss, M.R., Foulkes, S.B.} 
         {2007} {\MNRAS} {378} {785}

\refitem {Uemura, M. {\it et al.}} {2004} {\em Publ.Astr.Soc.Japan} {56} {141}

\refitem {Warner, B.} {2003} { {\it Cataclysmic Variable Stars}, 
                      {\rm 2nd edition, Cambridge University Press }} {~} {~}

\refitem {Whitehurst, R.} {1988} {\MNRAS} {232} {35} 

\refitem {Zhang, E.-H., Robinson, E.L., Nather, R.E.} {1986} {\ApJ} {305} {740}

\end {references}

\end{document}